\documentclass[useAMS,usenatbib]{mn2e}

\usepackage{amssymb}
\usepackage{amsmath}

\newcommand{\lya}{Ly$\alpha$}
\newcommand{\HI}{H\,{\sc i}}

\newcommand{\nhi}{\mbox{$N_{\rm HI}$}}
\newcommand{\Lya}{\mbox{Ly$\alpha$}}
\newcommand{\msun}{\mbox{$M_\odot$}}
\newcommand{\mhi}{\mbox{$M_{\rm HI}$}}
\newcommand{\kms}{\mbox{km s$^{-1}$}}
\newcommand{\cm}{cm$^{-2}$}

\newcommand{\dndz}{$dN_{DLA}/dz$}

\newcommand{\degree}{\hbox{$^\circ$}}

\newcommand{\wfi}{$w_{\rm 50}$}

\title[DLA Kinematics from 21-cm Measurements]{Evolution of damped Lyman $\alpha$ 
kinematics and the effect of spatial resolution on 21-cm measurements} 
\author[E.V. Ryan-Weber]{Emma V. Ryan-Weber,$^{1,2}$\thanks{email:
	eryan@ast.cam.ac.uk} Lister Staveley-Smith$^3$,
  Rachel L. Webster$^2$\\
  $^1$Institute of Astronomy, University of Cambridge\\
  $^2$School of Physics, Univeristy of Melbourne, VIC 3010 Australia\\
  $^3$Australia Telescope National Facility, CSIRO, PO Box 76, Epping,
  NSW 1710, Australia}

\begin{document}

\date{Accepted ***. Received 2005 September 5; in original form 2005 August 19}

\pagerange{\pageref{firstpage}--\pageref{lastpage}} \pubyear{2005}

\maketitle
 
\label{firstpage}

\begin{abstract}
We have investigated the effect of spatial resolution on determining
pencil-beam like velocity widths and column densities in
galaxies. Three 21-cm datasets are used, the HIPASS galaxy catalogue,
a subset of HIPASS galaxies with ATCA maps and a high-resolution image
of the LMC. Velocity widths measured from 21-cm emission in local
galaxies are compared with those measured in intermediate redshift
Damped Lyman-$\alpha$ (DLA) absorbers. We conclude that spatial
resolution has a severe effect on measuring pencil-beam like velocity
widths in galaxies. Spatial smoothing by a factor of 240 is shown to
increase the median velocity width by a factor of two. Thus any
difference between velocity widths measured from global profiles or
low spatial resolution 21-cm maps at $z=0$ and DLAs at $z>1$ cannot
unambiguously be attributed to galaxy evolution. The effect on column
density measurements is less severe and the values of \dndz\ from local
low-resolution 21-cm measurements are expected to be overestimated by
only $\sim$10 per cent.

\end{abstract}


\section{Introduction}

Damped Lyman-$\alpha$ (DLA) absorption systems provide the most
practical measure of neutral gas in intermediate redshift
galaxies. However, measurements of DLAs provide information in one
dimension only -- the redshift axis, and only along a single
pencil-beam line-of-sight. In rare cases of close binary QSO pairs or
gravitationally lensed QSOs, spatial information on the absorbers can
be deduced \citep[e.g.][]{Rauch02, Ellison04}. The nature of DLAs can
be inferred by comparing observations of DLA velocity widths to models
of galaxy evolution.

DLA kinematics and models of galaxy evolution were first compared by
\cite{Prochaska97} who concluded that rapidly rotating protogalactic
discs provide the best fit to DLA kinematics. If DLAs are the
progenitors of spiral discs, then their velocity widths are expected
to match that of present day galaxies. This interpretation however
seems to run contrary to cold dark matter theory.

The current structure formation paradigm predicts that star formation
first occurs in subgalactic clumps housed in the most massive dark
matter halos. This framework is well matched by the assembly of stars
and dark matter into galaxies, where more massive halos transform a
larger fraction of their gas into stars at high
redshift \citep{Jimenez05}. Further evidence is provided by the
clustering strengths of Lyman break galaxies (LBGs) at $z\sim3$,
showing that LBGs are embedded in massive dark matter halos
\citep{Adelberger05}. DLAs on the other hand are quite distinct from
LBGs \citep{Hopkins05}, perhaps corresponding to lower-mass systems
with lower rotational velocities. Since lower-mass systems convert
their gas into stars at a slower rate they contribute significantly to
the total DLA cross-section (averaged over time).

Hydrodynamic simulations demonstrate that irregular protogalactic
clumps can equally reproduce the observed velocity width distribution
of DLAs through a combination of rotation, random motions, infall and
merging \citep{Haehnelt98}. Both semi-analytic and numerical models
show that the average velocity width of DLAs is expected to increase
with evolution to $z=0$ \citep{Cen03,Okoshi04,Nagamine05}.  Other work
\citep{Maller01} interprets the kinematics of DLAs as evidence that
they arise in galaxies undergoing tidal stripping and mergers.


Because very few DLAs are known at low redshift, it is difficult to
measure their statistics directly.  Fortunately, at $z=0$ the
alternative method of 21-cm emission can be used. It has the added
advantage of known galaxy properties so the distribution of DLA galaxy
properties such as luminosity and \HI\ mass can be
inferred. \cite{Rosenberg03} have compared global 21-cm velocity
widths to velocity widths of intermediate redshift DLAs. They conclude
that evolution is not the likely cause of the difference between the
two populations since velocity widths in their sample of local
galaxies are as large or larger than the DLA velocity widths. Instead
they suggest that the difference between the two populations could be
caused by small number statistics or differences in the way the two
populations were measured. Unfortunately 21-cm emission measurements
always come with the caveat of spatial resolution as the typical beam
size is much larger than the region probed by DLA observations. The
extent of the absorbing region in DLAs is only as large as the
projected angular extent of the UV-emission region of the background
QSO. An upper limit to the optical emission region of one QSO is known
to be $\lesssim$0.03 pc by microlensing \citep{Wyithe02}. The
effective angular extent of absorption in each DLAs is expected to be
roughly constant with redshift as the angular diameter distance varies
less than 20 per cent over $1<z<4$, where most QSO probes and DLAs are
detected.

In this paper the relationship between \HI\ velocity widths in local
galaxies and intermediate redshift DLAs is investigated. Does
the velocity width distribution of neutral gas in galaxies change with
evolution to $z=0$ ? In addressing this question, the relationship of
pencil-beam DLA measurements to the global \HI\ profile of galaxies
and the effect of spatial resolution on determining \HI\ column
densities and velocity widths in 21-cm maps must also be investigated.

\section{DLA Kinematics}

Here we test the evolution of galaxy kinematics by comparing DLAs at
intermediate redshifts with gas-rich galaxies at the present epoch.
Velocity widths from two different 21-cm datasets are utilised: the
4315 galaxies in HICAT, which is the \HI\ Parkes All Sky Survey
(HIPASS) galaxy catalogue \citep{Meyer04} and a subset of 34 HICAT
galaxies with Australia Telescope Compact Array (ATCA) maps
\citep{Ryan-Weber03}. Velocity widths from these 21-cm observations
are compared with a complilation of velocity widths from 94 DLAs
\citep{Wolfe05}. The velocity widths of DLAs are measured using low
ionisation metal lines such as Si$^{+}$, Fe$^{+}$ and Ni$^{+}$ to
trace the neutral gas, as the \Lya\ line, by definition, is saturated
and does not resolve the DLA kinematics. Velocity profiles of low
ionisation metal lines trace each other well and therefore are assumed
to trace the kinematics of the cold gas. This assumption could be
verified with high resolution IUE spectra towards SN 1987A
\citep{Blades88}.

\begin{figure}
\vspace{14pc}
\includegraphics{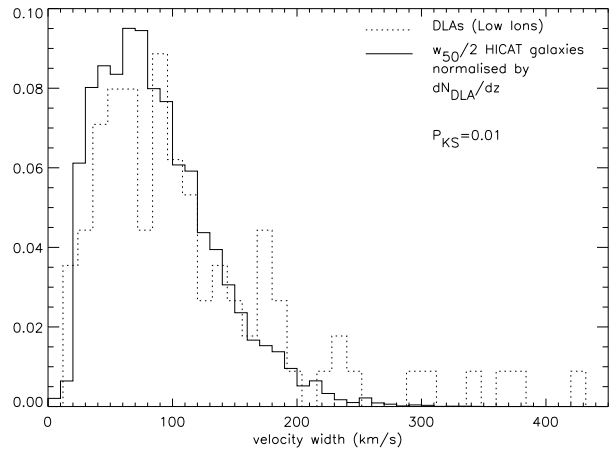}
\caption{Velocity width histogram comparing \wfi/2 from HICAT galaxies to
low ionisation metal lines in DLAs.}
\label{fig:w50DLAhicat}
\end{figure}

The global \HI\ profile statistic used from HICAT is \wfi, the
velocity width at 50 per cent of the profile maximum. Comparisions are
made with \wfi/2 as a single sight-line though a galaxy will
presumably only sample one half of a galaxy's rotation. The relative
contribution of each galaxy has been weighted by the \HI\ mass
dependent DLA cross-section
\citep{Ryan-Weber03}. Figure~\ref{fig:w50DLAhicat} shows the
distribution of \wfi/2 and velocity widths of low ionisation lines in
DLAs. A Kolmogorov-Smirnov (K-S) test shows the probability that the
two data sets are drawn from the same distribution is $P_{KS}= 0.01$,
compared with \cite{Rosenberg03}, who find $P_{KS}= 0.89$ and conclude
that evolution in the kinematics of \HI-rich galaxies cannot be ruled
out (a value of $P_{KS}\geq 0.99$ would rule out evolution). The same
conclusion is then reached with the HICAT dataset: evolution cannot be
ruled out, the two populations may differ due to evolution, but could
also differ due to small number statistics or differences in the way
the velocity widths are measured.


Differences in the way the velocity widths are measured can be reduced
by measuring the 21-cm widths with a method that most closely
resembles the determination of QSO absorption line velocity widths.
\cite{Prochaska97} measure the velocity width across the central 90
per cent of the optical depth. The ATCA 21-cm maps are used to measure
a {\it{pixel}} velocity width along the velocity axis of each spatial
pixel in each galaxy map. Only spatial pixels that satisfy the DLA
cutoff, \nhi$\geq2\times10^{20}$ \cm, are included. The velocity width
is determined for the central 90 per cent of the integrated profile,
this measurement is denoted by $\Delta v_{90\%}$. The average restored
beam of the ATCA maps is 1\arcmin, corresponding to a physical extent
of 2 to 7 kpc.

\begin{figure}
\vspace{14pc}
\includegraphics{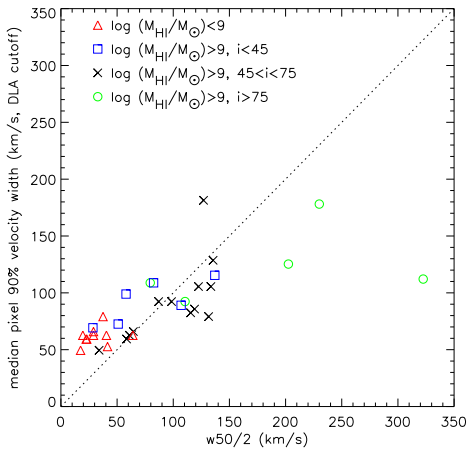}
\caption{Global \wfi/2 velocity width versus median $\Delta v_{90\%}$
for ATCA galaxies. The dotted line is where the two axes are equal.}
\label{fig:kin_med}
\end{figure}

Figure~\ref{fig:kin_med} shows that neither \wfi\ nor \wfi/2 provides
a good statistical representation of the median $\Delta v_{90\%}$
velocity width. Only 18 of the 34 galaxies in the sample have \wfi/2
measurements that fall within 30 \kms\ of their median $\Delta
v_{90\%}$. This is not a result of the DLA nor 90 per cent cutoff, as
a similar spread of median velocity widths is found when all spatial
pixels with \nhi$\geq10^{19}$ \cm\ or 100 per cent of the integrated
profile is considered.


The reason why \wfi\ widths are not well represented by pixel velocity
widths is due to the variety of galaxy types in the sample. This is
problematic as it is now well established that the local galaxy
population weighted by DLA \HI\ cross section samples a variety of
galaxy types \citep{Ryan-Weber03, Zwaan05}. Low mass and dwarf
galaxies tend to have Gaussian velocity profiles that remain
approximately constant over the area of the 21-cm map. The $\Delta
v_{90\%}$ method acting on a Gaussian profile, will return velocity
widths which are by definition 29 per cent larger than the FWHM, that
is, $\sim60$ per cent larger than \wfi/2. Face-on spiral galaxies are
expected to follow a similar overestimate of velocity widths, but not
as extreme as the low mass galaxies. On the other hand, the pixel
velocity widths of large edge-on spiral galaxies that have many pixels
spread over a large area are more likely to be lower than
\wfi/2. These trends are summarised in Figure~\ref{fig:kin_med}. As
expected, all the low mass (log \mhi/\msun $<9$) galaxies lie on or
above the line where \wfi/2 equals the median $\Delta v_{90\%}$. Most
face-on (inclination $<45\degree$) high mass (log \mhi/\msun $>9$)
galaxies also lie on the same side of the line. However, most massive
edge-on (inclination $>75\degree$) galaxies have median $\Delta
v_{90\%}$ that are well below their global \wfi/2 measurements. The
range of galaxy types means than {\it{any}} statistic of the pixel
velocity width, e.g. some fraction of the maximum velocity width, will
not be well represented by any global velocity width statistic. We
conclude that the global \wfi\ value does not provide a good estimate
of pencil-beam velocity widths, as measured in DLAs.


\begin{figure}
\vspace{14pc}
\includegraphics{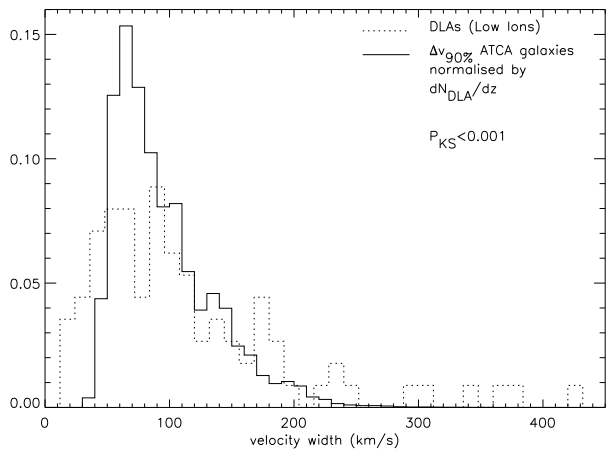}
\caption{Histogram comparing $\Delta v_{90\%}$ from the ATCA 
galaxy sample with DLA velocity widths.}
\label{fig:dv90DLAhist}
\end{figure}

Pixel velocity widths from the ATCA galaxies are compared with those
from DLAs in Figure~\ref{fig:dv90DLAhist}. The K-S probability is
calculated to be less than 0.001, indicating that the two
distributions are different. Does this mean that there is genuine
evolution in velocity width from the intermediate redshift DLAs to
present day \HI-rich galaxies, or is the difference still due to the
measurement techniques? The lack of high velocity widths could be
caused by treating galaxies as individual systems, if their satellites
were considered, larger velocity widths could be invoked. A deficit of
low velocity widths cannot be caused by velocity resolution since the
resolution is 3.3 \kms\ in the ATCA images. The galaxies that tend to
have the lowest velocity widths also have the smallest contribution to
\dndz, however there seems to be a complete lack of $\Delta v_{90\%}$
values below 40 \kms. The DLA column density cutoff is well above the
noise limit in the ATCA images, so the signal-to-noise ratio cannot be
an issue. Discrepancies between the velocity width of gas measured in
emission and absorption would ideally be tested with UV spectroscopy
towards QSOs recently identified behind the Large Magellanic Cloud
\citep[LMC,][]{Geha03, Dobrzycki05}, an experiment that awaits the
next space-based UV spectrograph.

\section{Velocity Width Statistics with the LMC}
\label{sect:vel_lmc}

The absence of low velocity line widths could be caused by the
relatively low spatial resolution of the ATCA galaxy maps. Following
on from \cite{Prochaska02} a comparison can be made using the \HI\
data cube of LMC. The LMC data cube is a high-resolution 21-cm data
cube constructed from the combined Parkes and ATCA survey of
\cite{Kim03}. The data has a velocity resolution of 1.6 \kms\ and a
column density sensitivity of 4$\times10^{19}$ \cm\ (for a line width
of 40 \kms).  The LMC has an \HI\ mass of 3.8$\times10^{8}$\msun, a
global \HI\ velocity width, \wfi=80 \kms, is located at a distance of
50 kpc \citep{Staveley-Smith03} and has an inclination of 42\degree\
\citep{Weinberg01}. The ATCA galaxy and LMC maps both have an angular
resolution of $\sim$1\arcmin. However, the LMC is 80 to 480 times
closer than the galaxies in the ATCA sample, leading to a large
difference in physical spatial resolution, $2-7$ kpc per beam for the
ATCA galaxy sample compared with 15 pc per beam for the LMC.

\begin{figure}
\vspace{14pc}
\includegraphics{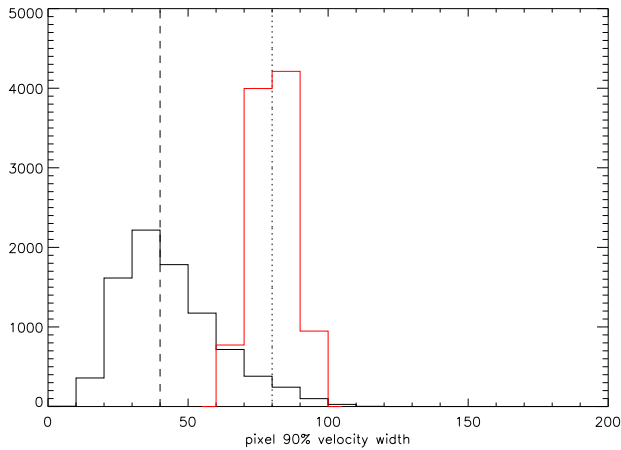}
\caption{Histogram of $\Delta v_{90\%}$ produced from $10^4$ random
pixels in the LMC cube satisfying \nhi\ $>2\times10^{20}$ \cm. The red
histogram is the same calculation for the LMC cube convolved with a
circular Gaussian to produce a smoothed image with FWHM =
4\degree. The dotted and dashed lines are \wfi\ and \wfi/2
respectively.}
\label{fig:dvLMC}
\end{figure}

Since the LMC data cube contains over $4\times10^{6}$ spatial pixels,
a random sample of $10^4$ pixels is used in this analysis. The
$\Delta v_{90\%}$ method was applied to this random sample of $10^4$
pixels and the results are given in Figure~\ref{fig:dvLMC}. Only
pixels that satisfy the DLA cutoff, \nhi~$\geq2\times10^{20}$\cm, were
included. The LMC cube was convolved with a 2-D circular Gaussian in
the spatial plane. The LMC cube was smoothed on three scales such that
the resolution in the smoothed cubes were 80, 240 and 480 times
greater than the original. A resolution factor of 240 corresponds to a
final resolution of 4\degree, effectively placing the LMC at 12 Mpc,
or the maximum distance at which galaxies with \HI\ masses similar to
the LMC are found in the ATCA galaxy sample.

The velocity width histogram for this smoothed cube is plotted in
Figure~\ref{fig:dvLMC}. The difference is stark and shows that the
median velocity width from the smoothed cube using the $\Delta
v_{90\%}$ method is now comparable to \wfi, rather than \wfi/2.
Decreasing the resolution shows an overall increase in the median
velocity width is because the velocity width of the smoothed set of
pixels will return the maximum velocity width rather than the average
velocity width. The median velocity width continually increases with
decreasing resolution: the unsmoothed cube with a resolution of
1\arcmin\ and smoothed cubes with resolutions 1.3\degree, 4\degree\
and 8\degree\ have median velocity widths of 41, 64, 81 and 87 \kms\
respectively. Turning this statement around, as the resolution
increases, the distribution of velocity width spreads from half the
low resolution minimum velocity width to the low resolution maximum
velocity width and the median is reduced by half.  Applying this
general trend to the ATCA galaxy sample in
Figure~\ref{fig:dv90DLAhist}, a distribution more like the DLAs is
obtained. Therefore this resolution effect cannot be decoupled from
evolution in the velocity width distribution between local gas-rich
galaxies and higher redshift DLAs.

The highest resolution LMC cube probes structure on scales 500 times
greater than that of DLA measurements. Decreasing the spatial
resolution in the LMC cube a further 240 times essentially doubles the
median velocity width. If \HI\ structure is self-similar on smaller
scales, then one would expect the velocity widths to decrease further
as the scale on which DLAs are measured is approached.

\section{The Effect of Spatial Resolution on Column Density}
\label{sect:resnhi}

Since spatial resolution was found to have such a dramatic effect on
velocity width measurements, what is its effect on the other 21-cm
based quantities? There is only one case of \HI\ emission detected
from a DLA galaxy. The DLA galaxy SBS 1543+593 \citep{Bowen01a} has
been mapped using GMRT \citep{Chengalur02} in 21-cm. The column
density measured at the QSO location in the GMRT map is
$5\times10^{20}$\cm, in reasonable agreement with the \lya\ column
density of $2.2\times10^{20}$\cm. Here the effect of resolution on the
column density distribution function, $f(\nhi)$, and the cross-section
of \HI\ satisfying the DLA criteria, that is the number density of
DLAs per unit redshift, \dndz, is investigated.

\begin{figure}
\vspace{14pc}
\includegraphics{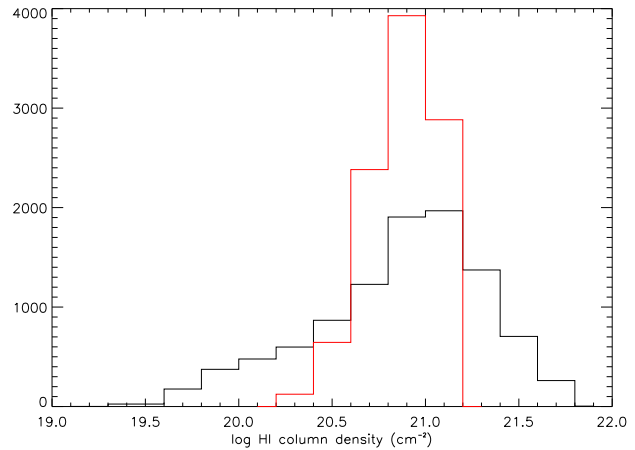}
\caption{Histogram of log column density from $10^4$ random pixels
in the LMC cube. The red histogram is of the same smoothed image
described in Figure~\ref{fig:dvLMC}. No \nhi\ restrictions are applied
to either data set.}
\label{fig:LMC_nhi}
\end{figure}

The column density histogram for $10^4$ random spatial pixels in the
LMC cube is given in Figure~\ref{fig:LMC_nhi} together with the
histogram derived from the LMC cube convolved with a circular Gaussian
to produce a cube smoothed in the spatial plane with FWHM = 4\degree.
The full range of column densities are considered here. Lowering the
resolution in the cube means that any fine structure in low or high
column density regions is washed out. The histogram reflects this
qualitative expectation and shows that the distribution of column
densities are truncated at both the low and high ends.


\begin{figure}
\vspace{14pc}
\includegraphics{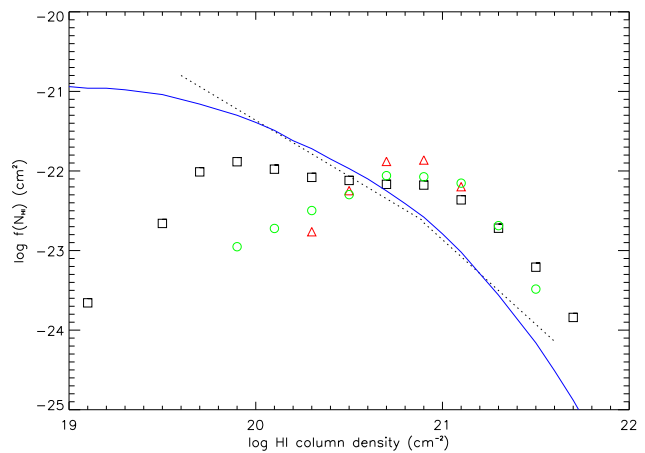}
\caption{LMC $f(\nhi)$ (black squares), with arbitrary normalisation. 
Green circles and red triangles give the LMC $f(\nhi)$ 
for smoothed cubes with FWHM =
1.3\degree\ and 4\degree\ respectively. The dotted line is the 
double power law fit from Ryan-Weber et al. (2003), with normalisation 
correction from Ryan-Weber et al. (2005). The solid blue line is 
$f(\nhi)$ from Zwaan et al. (2005).}
\label{fig:LMC_cddf}
\end{figure}

The column density distribution function will naturally be influenced
by this change in distribution. Figure~\ref{fig:LMC_cddf} shows
$f(\nhi)$ from \cite{Ryan-Weber03} and \cite{Zwaan05} together with
$f(\nhi)$ calculated from $10^4$ random spatial pixels in the LMC cube
at the full resolution and smoothed at two different levels.  Since
only one galaxy, the LMC, is considered, the same arbitrary
normalisation is applied to all the LMC $f(\nhi)$ points. None of the
LMC $f(\nhi)$ functions resemble the shape of $f(\nhi)$ calculated
from many galaxies. The shape of $f(\nhi)$ does vary between galaxies
and a turnover at \nhi $\sim10^{20}$ \cm\ occurs in $\sim$ 10 per cent
of the ATCA galaxies (see Ryan-Weber et al. (2003) figure A2). The
plot shows that decreasing the resolution creates a hump in the centre
of the function. Applying this trend to $f(\nhi)$ calculated from the
lower resolution ATCA maps, suggests that the `real' underlying
$f(\nhi)$ could be described by a single power law, as the lower \nhi\
slope would steepen and the upper \nhi\ slope would flatten. Although
this change in shape is based on one galaxy only.

The calculation of the number of DLAs per unit redshift at $z=0$,
\dndz, is also affected by resolution. The DLA cross-section of the
LMC increases slightly with decreasing resolution: the increase in the
DLA cross-section from the original resolution to cubes smoothed to a
resolution of 1.3, 4 and 8\degree\ is 11, 12, and 12 per cent
respectively. Hence the overall effect of resolution on local \dndz\
calculations \citep{Rosenberg03,Ryan-Weber03, Ryan-Weber05,Zwaan05} is
small, and this $11-12$ per cent increase lies within the
uncertainties for the measured values of \dndz.

\section{Summary}

The results highlight the difference in conclusions when velocity
widths are measured in different ways. The \wfi\ velocity width from a
galaxy's global \HI\ profile is not well represented by the
distribution of pixel (pencil-beam like) velocity widths in a
galaxy. Instead, the relationship between \wfi\ and median pixel
velocity width was found to depend on galaxy mass and
inclination. This work provides a solid basis for the suggestion by
\cite{Rosenberg03} that evolution in DLA kinematics (measured by 21-cm
global velocity widths at $z=0$, and absorption line profiles at
higher redshift) cannot be decoupled from the different measurement
techniques. Thus using DLA velocity widths to infer the total velocity
width of a galaxy and related quantities, e.g. mass, must also be used
with caution \citep[e.g.][]{Godfrey05}.

Furthermore, when we use a method that most closely resembles that of
DLA velocity width measurements, the $\Delta v_{90\%}$ method, the
same conclusion is reached. The distribution of velocity widths
measured in DLAs at intermediate redshifts is different from those
measured along the line-of-sight of individual spatial pixels within
galaxies. It is not possible to decouple the effect of spatial
resolution in low-resolution 21-cm maps from evolution in galaxy
kinematics. Unfortunately the large difference in spatial resolution
sampled by 21-cm emission and \lya\ absorption cannot be
reconciled. The results of \cite{Prochaska02} still hold, as a single
low-mass galaxy such as the LMC cannot account for high ($>100$ \kms)
DLA velocity widths. However differences still exist between the
highest resolution 21-cm data, which have multi-component profiles
overlaping in velocity coverage, and DLA profiles, which tend to
decompose into several components distinct in velocity coverage
\citep[e.g.][]{Wolfe05}.

Column density measurements and related calculations are not adversely
affected by spatial resolution. Lowering the resolution in 21-cm data
means that fine structure in low or high column density regions is
washed out. This causes a central hump in the column density
distribution function. The expected increase due to spatial resolution
effects in calculating \dndz\ at z=0 is within uncertainties quoted in
\cite{Rosenberg03}, \cite{Ryan-Weber03,Ryan-Weber05} and
\cite{Zwaan05}.

\section*{Acknowledgments}

EVR-W acknowledges support from the University of Melbourne 
A.J. Shimmins Postgraduate Writing-Up Award.

\bibliographystyle{mn2e}
\bibliography{mn-jour,res}

\begin{thebibliography}{}

\bibitem[\protect\citeauthoryear{{Adelberger}, {Steidel}, {Pettini}, {Shapley},
  {Reddy} \& {Erb}}{{Adelberger} et~al.}{2005}]{Adelberger05}
{Adelberger} K.~L.,  {Steidel} C.~C.,  {Pettini} M.,  {Shapley} A.~E.,  {Reddy}
  N.~A.,    {Erb} D.~K.,  2005, ApJ, 619, 697

\bibitem[\protect\citeauthoryear{{Blades}, {Wheatley}, {Panagia}, {Grewing},
  {Pettini} \& {Wamsteker}}{{Blades} et~al.}{1988}]{Blades88}
{Blades} J.~C.,  {Wheatley} J.~M.,  {Panagia} N.,  {Grewing} M.,  {Pettini} M.,
     {Wamsteker} W.,  1988, ApJ, 334, 308

\bibitem[\protect\citeauthoryear{{Bowen}, {Tripp} \& {Jenkins}}{{Bowen}
  et~al.}{2001}]{Bowen01a}
{Bowen} D.~V.,  {Tripp} T.~M.,    {Jenkins} E.~B.,  2001, AJ, 121, 1456

\bibitem[\protect\citeauthoryear{{Cen}, {Ostriker}, {Prochaska} \&
  {Wolfe}}{{Cen} et~al.}{2003}]{Cen03}
{Cen} R.,  {Ostriker} J.~P.,  {Prochaska} J.~X.,    {Wolfe} A.~M.,  2003, ApJ,
  598, 741

\bibitem[\protect\citeauthoryear{{Chengalur} \& {Kanekar}}{{Chengalur} \&
  {Kanekar}}{2002}]{Chengalur02}
{Chengalur} J.~N.,  {Kanekar} N.,  2002, A\&A, 388, 383

\bibitem[\protect\citeauthoryear{{Dobrzycki}, {Eyer}, {Stanek} \&
  {Macri}}{{Dobrzycki} et~al.}{2005}]{Dobrzycki05}
{Dobrzycki} A.,  {Eyer} L.,  {Stanek} K.~Z.,    {Macri} L.~M.,  2005, A\&A,
  442, 495

\bibitem[\protect\citeauthoryear{{Ellison}, {Ibata}, {Pettini}, {Lewis},
  {Aracil}, {Petitjean} \& {Srianand}}{{Ellison} et~al.}{2004}]{Ellison04}
{Ellison} S.~L.,  {Ibata} R.,  {Pettini} M.,  {Lewis} G.~F.,  {Aracil} B.,
  {Petitjean} P.,    {Srianand} R.,  2004, A\&A, 414, 79

\bibitem[\protect\citeauthoryear{{Geha et al.}}{{Geha et al.}}{2003}]{Geha03}
{Geha et al.} 2003, AJ, 125, 1

\bibitem[\protect\citeauthoryear{{Godfrey} \& {Francis}}{{Godfrey} \&
  {Francis}}{2005}]{Godfrey05}
{Godfrey} L.~E.~H.,  {Francis} P.~J.,  2005, Publications of the Astronomical
  Society of Australia, 22, 245

\bibitem[\protect\citeauthoryear{{Haehnelt}, {Steinmetz} \& {Rauch}}{{Haehnelt}
  et~al.}{1998}]{Haehnelt98}
{Haehnelt} M.~G.,  {Steinmetz} M.,    {Rauch} M.,  1998, ApJ, 495, 647

\bibitem[\protect\citeauthoryear{{Hopkins}, {Rao} \& {Turnshek}}{{Hopkins}
  et~al.}{2005}]{Hopkins05}
{Hopkins} A.~M.,  {Rao} S.~M.,    {Turnshek} D.~A.,  2005, ApJ, 630, 108

\bibitem[\protect\citeauthoryear{{Jimenez}, {Panter}, {Heavens} \&
  {Verde}}{{Jimenez} et~al.}{2005}]{Jimenez05}
{Jimenez} R.,  {Panter} B.,  {Heavens} A.~F.,    {Verde} L.,  2005, MNRAS, 356,
  495

\bibitem[\protect\citeauthoryear{{Kim}, {Staveley-Smith}, {Dopita}, {Sault},
  {Freeman}, {Lee} \& {Chu}}{{Kim} et~al.}{2003}]{Kim03}
{Kim} S.,  {Staveley-Smith} L.,  {Dopita} M.~A.,  {Sault} R.~J.,  {Freeman}
  K.~C.,  {Lee} Y.,    {Chu} Y.-H.,  2003, ApJS, 148, 473

\bibitem[\protect\citeauthoryear{{Maller}, {Prochaska}, {Somerville} \&
  {Primack}}{{Maller} et~al.}{2001}]{Maller01}
{Maller} A.~H.,  {Prochaska} J.~X.,  {Somerville} R.~S.,    {Primack} J.~R.,
  2001, MNRAS, 326, 1475

\bibitem[\protect\citeauthoryear{{Meyer et al.}}{{Meyer et
  al.}}{2004}]{Meyer04}
{Meyer et al.} 2004, MNRAS, 350, 1195

\bibitem[\protect\citeauthoryear{{Nagamine}, {Springel} \&
  {Hernquist}}{{Nagamine} et~al.}{2004}]{Nagamine05}
{Nagamine} K.,  {Springel} V.,    {Hernquist} L.,  2004, MNRAS, 348, 421

\bibitem[\protect\citeauthoryear{{Okoshi}, {Nagashima}, {Gouda} \&
  {Yoshioka}}{{Okoshi} et~al.}{2004}]{Okoshi04}
{Okoshi} K.,  {Nagashima} M.,  {Gouda} N.,    {Yoshioka} S.,  2004, ApJ, 603,
  12

\bibitem[\protect\citeauthoryear{{Prochaska}, {Ryan-Weber} \&
  {Staveley-Smith}}{{Prochaska} et~al.}{2002}]{Prochaska02}
{Prochaska} J.~X.,  {Ryan-Weber} E.,    {Staveley-Smith} L.,  2002, PASP, 114,
  1197

\bibitem[\protect\citeauthoryear{{Prochaska} \& {Wolfe}}{{Prochaska} \&
  {Wolfe}}{1997}]{Prochaska97}
{Prochaska} J.~X.,  {Wolfe} A.~M.,  1997, ApJ, 487, 73

\bibitem[\protect\citeauthoryear{{Rauch}, {Sargent}, {Barlow} \&
  {Simcoe}}{{Rauch} et~al.}{2002}]{Rauch02}
{Rauch} M.,  {Sargent} W.~L.~W.,  {Barlow} T.~A.,    {Simcoe} R.~A.,  2002,
  ApJ, 576, 45

\bibitem[\protect\citeauthoryear{{Rosenberg} \& {Schneider}}{{Rosenberg} \&
  {Schneider}}{2003}]{Rosenberg03}
{Rosenberg} J.~L.,  {Schneider} S.~E.,  2003, ApJ, 585, 256

\bibitem[\protect\citeauthoryear{{Ryan-Weber}, {Webster} \&
  {Staveley-Smith}}{{Ryan-Weber} et~al.}{2003}]{Ryan-Weber03}
{Ryan-Weber} E.~V.,  {Webster} R.~L.,    {Staveley-Smith} L.,  2003, MNRAS,
  343, 1195

\bibitem[\protect\citeauthoryear{{Ryan-Weber}, {Webster} \&
  {Staveley-Smith}}{{Ryan-Weber} et~al.}{2005}]{Ryan-Weber05}
{Ryan-Weber} E.~V.,  {Webster} R.~L.,    {Staveley-Smith} L.,  2005, MNRAS,
  356, 1600

\bibitem[\protect\citeauthoryear{{Staveley-Smith}, {Kim}, {Calabretta},
  {Haynes} \& {Kesteven}}{{Staveley-Smith} et~al.}{2003}]{Staveley-Smith03}
{Staveley-Smith} L.,  {Kim} S.,  {Calabretta} M.~R.,  {Haynes} R.~F.,
  {Kesteven} M.~J.,  2003, MNRAS, 339, 87

\bibitem[\protect\citeauthoryear{{Weinberg} \& {Nikolaev}}{{Weinberg} \&
  {Nikolaev}}{2001}]{Weinberg01}
{Weinberg} M.~D.,  {Nikolaev} S.,  2001, ApJ, 548, 712

\bibitem[\protect\citeauthoryear{{Wolfe}, {Gawiser} \& {Prochaska}}{{Wolfe}
  et~al.}{2005}]{Wolfe05}
{Wolfe} A.~M.,  {Gawiser} E.,    {Prochaska} J.~X.,  2005, ARA\&A, 43, 861

\bibitem[\protect\citeauthoryear{{Wyithe}, {Agol} \& {Fluke}}{{Wyithe}
  et~al.}{2002}]{Wyithe02}
{Wyithe} J.~S.~B.,  {Agol} E.,    {Fluke} C.~J.,  2002, MNRAS, 331, 1041

\bibitem[\protect\citeauthoryear{{Zwaan}, {van der Hulst}, {Briggs},
  {Verheijen} \& {Ryan-Weber}}{{Zwaan} et~al.}{2005}]{Zwaan05}
{Zwaan} M.~A.,  {van der Hulst} J.~M.,  {Briggs} F.~H.,  {Verheijen} M.~A.~W.,
    {Ryan-Weber} E.~V.,  2005, MNRAS, accepted, astro-ph/0510127

\end{thebibliography}

\bsp

\label{lastpage}

\end{document}